\documentclass[prl,twocolumn]{revtex4}

\usepackage{graphicx}

\begin{document}

\title{Fermion pairing with population imbalance:\\
energy landscape and phase separation in a constrained Hilbert
subspace}
\author{Zheng-Cheng Gu$^{\dagger,\dagger\dagger}$, Geoff Warner$^{\dagger}$ and Fei Zhou$^{\dagger}$}
\affiliation{ \textit{Department of Physics and Astronomy, The
University of British Columbia, Vancouver, B.C., Canada
V6T1Z1$^{\dagger}$}\\
\textit{Center for Advanced Study, Tsinghua University, Beijing,
China, 100084$^{\dagger\dagger}$}}
\date{{\small \today}}

\begin{abstract}
In this Letter we map out the mean field energy potential landscape of fermion
pairing states with population imbalance near broad Feshbach
Resonances. We apply the landscape to investigate the nature of phase
separation, when the Hilbert space is subject to the constraint of constant population
imbalance. We calculate the scattering length dependence
of the critical population imbalance for various phase separated states across
Feshbach resonances.
\end{abstract}

\maketitle

Recently, cold atom pairing states with population differences were
studied experimentally by the MIT and the Rice University cold atom
groups\cite{Zwierlein,Partridge}. Fermion pairing with population
imbalance has been a fascinating subject for a long time and was
studied in various contexts
\cite{Clogston,Sarma,Fulde,Larkin,Liu,Bedaque}. For cold
atoms, pairing occurs between two hyperfine states participating in
Feshbach resonances, as indicated by experiments a few years
ago\cite{exp1}. Cold atom superfluids across Feshbach resonances
with population imbalance have also been studied theoretically, to some extent; different
phases were proposed, based on a variety of arguments or
calculations\cite{Carlson,Pao,Sheehy}. Most recently, cold atom
superfluids with population imbalance in finite traps were
addressed\cite{Duan06}.
It is generally believed that such superfluids exhibit different behaviors on either side of
resonances. In the BCS limit, where the
scattering length is negative and small in magnitude, the
superfluid undergoes a first-order phase transition to a LOFF state
(pairing between nonzero total momentum), followed by a second order
phase transition to a partially polarized normal state, as the
difference in chemical potentials between the two spin species is increased\cite{Fulde,Larkin}. 
On the other hand, on the far BEC side where the scattering length is 
positive and small, one has a continuous
transition to a superfluid state coexisting with a Fermi sea
composed of the excess spin species.

One of the unique features of cold atoms is that
relaxation processes
including the (global) population imbalance
relaxation are very slow compared to the experimental measurement time
period. Therefore, for cold atoms the population imbalance $P_{im}$
can be considered to be
{\em independent} of $I$, the energy splitting between two fermions with
the same momentum. 
This is opposed to traditional solid state,
many-body systems where the chemical potential difference between the two
species is a unique function of I, the Zeeman splitting energy, in equilibrium.
For cold atoms, the relevant ground state is therefore the
state of lowest energy in the constrained Hilbert space defined by fixed
total particle number N and population imbalance
$P_{im}$, not the lowest energy states in the whole Hilbert space
for a given total density. Motivated by the above observation and
the recent experiments on cold atom pairing, in this article we map
out the energy potential($\cal EP$) landscape of all lowest energy homogeneous states in
different subspaces specified by $P_{im}$.
(We leave out discussions on LOFF or other
more exotic competing translational-symmetry-breaking states in this article.)
The landscape not only provides important information on the
energetics of various states but is also critically valuable for
the understanding of phase separation phenomena (see below).
Furthermore, it paves the way for future studies of the
dynamics of superfluids in the limit of slow relaxation. As the first application, we
study the phase separation with the help of the landscape and
construct phase separated states at a given population imbalance
$P_{im}$ (and with a given density $\rho$). Finally we investigate critical
population imbalance for various phase separated states across Feshbach
resonances.

\begin{figure}[tbp]
\begin{center}
\includegraphics[width=3in]
{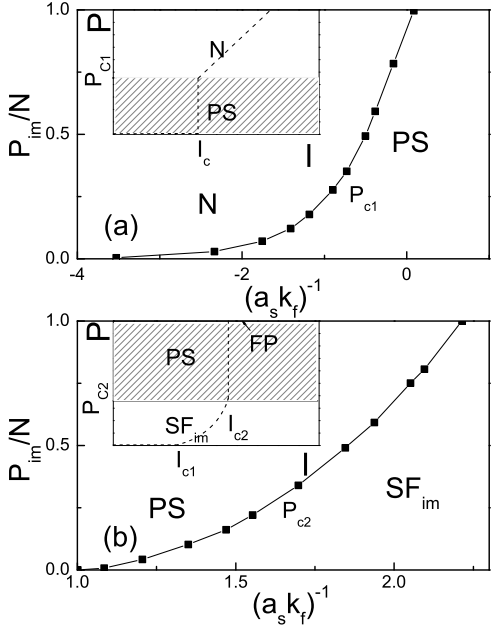}
\end{center}
\caption{The critical population imbalance as a function of
scattering length. In (a)[(b)], $P_{c1}(P_{c2})$ versus $(
a_sk_f)^{-1}$ for negative(positive) scattering length $a_s$.
$P_{c1}(P_{c2})$ is the critical value below(beyond) which phase
separation takes place(see also inset). Here $N$ stands for a normal
state, $PS$ for a phase separation state, $\text{SF}_{im}$ for a
superfluid with a finite {\em uniform} population imbalance, and FP for a
fully
polarized state. In (b), the phase separated state is constructed
using a superfluid with a finite population difference and a fully
polarized state. In inset, we show phases on the $P_{im}-I$ plane.
Shaded areas are for phase separated states; the dashed lines are
the schematic equilibrium curves $P_{im}=P^{eq}_{im}(I)$ for a given
chemical potential, towards which
the system is driven if an exchange of particles with a reservoir
is allowed.
\label{1}}
\end{figure}

The phase separation phenomenon discussed here only occurs
when we consider states in a Hilbert subspace with a conserved
population imbalance. These phase separated states generally do not
correspond to ground states in the whole Hilbert space for a fixed
total number of particles. In a population-imbalance-conserved subspace
phase separation occurs at any energy splitting $I$.
The main results are summarized in Fig.1. Unlike in
usual superconductors where a state prepared at an arbitrary
$P_{im}$ and $I$ generally is driven toward an equilibrium state
with $P_{im}=P_{im}^{eq}(I)$, a cold atom state can only settle with
the lowest energy state in the Hilbert subspace of a given initial
population imbalance $P_{im}$. Phases in the inset of Fig. \ref{1}
(a) and (b) are for those lowest energy states with a given
population difference $P_{im}$ and density $\rho$ in a given field $I$.
In Fig. \ref{1} (a) and (b) we present
the scattering length dependence of critical population imbalance
$P_{c1}, P_{c2}$.

\begin{figure}[tbp]
\begin{center}
\includegraphics[width=3.5in]
{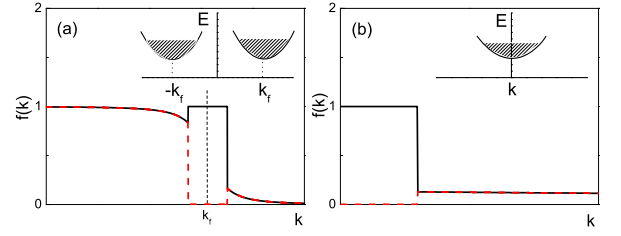}
\end{center}
\caption{The momentum distribution function $f(k)$ of spin-up(solid
line) and spin-down atoms(dashed line) in homogeneous states studied
in this article. (a) For negative scattering length and positive
chemical potential; (b) for positive scattering length and negative
chemical potential. Inset: schematic of the occupation of
quasi-particles.\label{3}}
\end{figure}

The model we employ to study this subject is the standard
one-channel model which is suitable for the discussion of
superfluids near broad Feshbach resonances. The homogeneous states
under consideration are BCS states with excess quasi-particles
accommodating unpaired atoms or the population imbalance, i. e.,
$\prod_{\textbf{k}\in \Gamma}\gamma_{\textbf{k},\uparrow}^\dagger
|BCS\rangle $ ($\gamma_{\textbf{k},\uparrow}^\dagger$ is the creation operator of
quasiparticles).
These states are analogues of the breached-pair states discussed
previously\cite{Liu}. Notice here $|BCS\rangle$ is a BCS state with
a $P_{im}$-dependent gap
  $\Delta(P_{im})$ and $\gamma_{\textbf{k},\uparrow}^\dagger$
  is the quasiparticle creation operator.
   The corresponding gap is self-consistently
determined by the following gap equation in the presence of
population imbalance
\begin{eqnarray}
\frac{1}{\lambda_R}={\sum_{\textbf{k}}}\frac{1}{2E_\textbf{k}}
(1-n_{\textbf{k},\uparrow})
-\sum_\textbf{k}\frac{1}{2\varepsilon_\textbf{k}}\label{rsf1}
\end{eqnarray}
with $\varepsilon_\textbf{k}=k^2/2m$,
$\xi_\textbf{k}=\varepsilon_\textbf{k}-\mu$,
$E_k=\sqrt{\Delta^2+\xi_\textbf{k}^2}$ and
$n_{\textbf{k},\uparrow}$ is the occupation number of quasi
particles at state k, spin-up and $\lambda_R$ is related to $a_s$ by
$\lambda_R V=-4\pi a_s/m$. Furthermore, only the lowest energy
quasi-particle states are occupied so as to minimize the kinetic energy
for given population imbalance. $n_{\textbf{k},\uparrow}$
therefore takes the form of a step function
\begin{equation}
n_{\textbf{k},\uparrow}=\left\{\begin{array}{ccc}1 & \rm{if}
&|\varepsilon_\textbf{k}-\mu|\leq \delta\xi\\0 &\rm{if}&
|\varepsilon_\textbf{k}-\mu|> \delta\xi\end{array}\right.
\end{equation}
which is unity only when $|\varepsilon_\textbf{k}-\mu|$ is less than
the cut-off energy $\delta\xi$. The energy of quasi-particles is
$E_k-I$ and can be either positive or negative; the distribution
function here is not the usual equilibrium one. The total number of
these quasiparticles precisely yields the population imbalance as
indicated in Fig. \ref{3},
\begin{equation}
P_{im}=\sum_\textbf{k}n_{\textbf{k},\uparrow} \label{ib}
\end{equation}
Each state under consideration here is therefore the lowest energy
homogeneous state in the corresponding constrained subspace. Finally,
the total number of particles is given as
\begin{eqnarray}
N={\sum_\textbf{k}}\left(1-\frac{\xi_\textbf{k}}{E_\textbf{k}}\right)
(1-n_{\textbf{k},\uparrow})
+\sum_\textbf{k}n_{\textbf{k},\uparrow}\label{rsf2}
\end{eqnarray}
The mean-field $\cal EP$, $\Omega=<{\cal H}_{BCS} >-\mu N$
(or the free energy at $T=0$ but with $P_{im}$ {\em fixed}) is
\begin{eqnarray}
\Omega={\sum_{\textbf{k}}}(E_\textbf{k}-
I)n_{\textbf{k},\uparrow}+{\sum_\textbf{k}}
 (\xi_\textbf{k}-E_\textbf{k}-\frac{\Delta^2}{2\varepsilon_\textbf{k}})+
\frac{\Delta^2}{\lambda_R} \label{renergy}
\end{eqnarray}
Taking into account Eq.(\ref{rsf2}) one can also easily obtain
the energies of states with given $N$ and $P_{im}$.

\begin{figure}[tbp]
\begin{center}
\includegraphics[width=3in]
{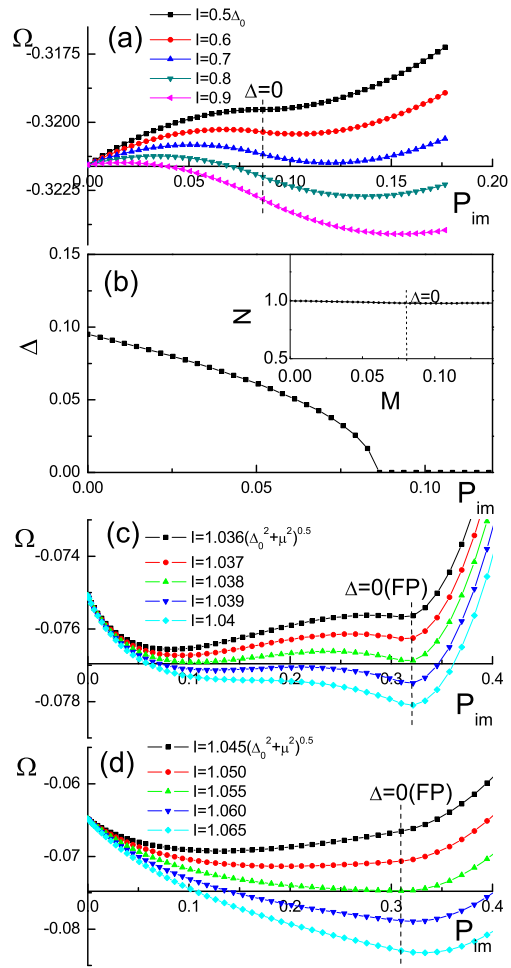}
\end{center}
\caption{In (a), (c) and (d) we show the energy potential as a function of the
population imbalance $P_{im}$ at different energy splitting $I$ but
with a fixed chemical potential. (a) is for a negative scattering
length $(a_s k_F^0)^{-1}=-1.40$ and $\mu=0.98\varepsilon_F^0$; (c)
for positive scattering length near resonance $(a_s
k_F^0)^{-1}=1.27$ and $\mu=-1.68\varepsilon_F^0$; (d) for a small
positive scattering length $(a_s k_F^0)^{-1}=1.56$ and
$\mu=-2.66\varepsilon_F^0$. $k_F^0$ is defined as the Fermi momentum
for the density $N_0/V$ at zero population imbalance $P_{im}=0$ and
$\varepsilon_F^0$ is the corresponding Fermi energy. In (b), we show
the variation of $\Delta$ and $N$ as a function of $P_{im}$ in the
homogeneous solutions to Eq. (1)-(4) (with the same scattering
length and chemical potential as in (a)). Notice in (c) and (d), all
the states to the right of the vertical dashed line correspond to
fully polarized states. Here $N$ and $P_{im}$ are expressed in unit
of $N_0$ and $\Omega$ is in unit of
$0.75^{2/3}\varepsilon_F^0$.\label{2}}
\end{figure}

The $\cal EP$ landscape of homogeneous states at different scattering length 
has been obtained. 
When the scattering length is negative and the magnitude is small,
we find a set of homogeneous solutions $\Delta(P_{im})$ to the gap
equation for a given population imbalance $P_{im}$. The
self-consistent BCS gap decreases as the magnetization $P_{im}$
increases and becomes zero at a critical value as seen in Fig.
\ref{2} (b). The critical value of $P_{im}$ is independent of the
energy splitting as also indicated by vertical lines in Fig. \ref{2}
(a), (c) and (e). The cold atoms in this limit would have a
continuous phase transition to a normal state if the spatial
phase-separation were prohibited.
The $\cal EP$ of states with a given chemical potential and in the presence of a given energy splitting
is plotted as a function of the population imbalance $P_{im}$ in Fig. \ref{2}.
Firstly, the BCS state represented by the solution at $P_{im}=0$
point becomes degenerate with a normal state (a solution with
$\Delta=0$) at a critical magnetic field $I_c=\Delta_0/\sqrt{2}$.
Secondly, a local minimum (a metastable normal state) and a maximum
(the Sarma solution) appear when the energy splitting is between
$I=\Delta_0/2$ and $I=\Delta_0$; here
the normal and BCS state are interpolated by the Sarma solution
as expected in the conventional BCS theory\cite{Sarma}.

To find the ground state for a given conserved population
imbalance, we examine the $\cal EP$ landscape of the homogeneous states.
We find that at {\em any given energy splitting} the energy curve is
concave (down) when the population imbalance is small. 
For a given population
difference, one constructs a mixed state which involves the BCS state and a normal 
state located at the end point of the concave part of the curve.
These two states involved in
the construction 
correspond to two degenerate states at the
first order phase transition point and 
always have the same chemical potentials for each fermion species.
For a given density $\rho$ and $P_{im}$,
$X$ the fraction of the BCS state and $\mu$ should be determined self-consistently
by the following conditions,

\begin{eqnarray}
&& \rho=\rho_{BCS}(\mu) X + \rho_N (\mu,\delta\mu=I_c(\mu))(1-X),\nonumber \\
&& P_{im}=P_{im}(\mu,\delta\mu=I_c(\mu))(1-X).
\end{eqnarray}
Here $\rho_{BCS(N)}$ is the density of a BCS (normal) state
when the chemical potential difference between two species $\delta \mu$ is equal to 
the critical energy splitting $I_c$;
$P_{im}$ is the corresponding population imbalance of the normal state.
The phase separation occurs when the population imbalance is smaller than a critical 
one.
The critical value corresponds to the population imbalance at the end point
of the concave-downwards part of the curve, or the solution to the above equation at $X=0$.
Above that critical
value the homogeneous states regain stability. By carrying out
similar analysis at different scattering length, we also obtain the
scattering length dependence of the critical population imbalance
(Fig. \ref{1}(a)).

When the scattering length becomes positive and especially when the
chemical potential becomes negative, the $\cal EP$ landscape
experiences an important qualitative change. The part of curve for
the small population imbalance becomes {\em convex} (or concave up) instead of
concave (down). On the other hand, beyond a critical imbalance population
the curve again is concave. A transition from {\em convex} to {\em
concave} as the population imbalance increases is a general feature
of the landscape when the chemical potential is negative and again
is independent of the energy splitting.

As the energy splitting increases, the global minimum along the
curve shifts {\em continuously} from $P_{im}=0$ point into a uniform
superfluid with a finite population imbalance. This signifies a
continuous transition from a BCS state to a uniform superfluid with
population imbalance when the chemical potential becomes negative. 
Further increasing the energy splitting
leads to an additional local minimum and local maximum (see Fig.
\ref{2} (c)). The new local minima located by the vertical dashed
line represent \emph{fully polarized states} and therefore the
pairing amplitude associated with these states is zero, i.e.
$\Delta=0$.
At an upper critical value, we obtain two degenerate
solutions located at two ends of the concave part of the curve, one
being a homogeneous superfluid with population imbalance and the
other a fully polarized state. 
One uses these two states 
to construct a phase separated state since again the chemical
potential for each fermion species can be shown to be the same.
Phase separation occurs when the population imbalance
exceeds a critical value specified by the left end of the 
concave curve in Fig.\ref{2}(c)
(i.e. the superfluid with uniform population imbalance).
Note that in this case the density variation along a given curve is substantial.
At a smaller positive scattering length or deep into the BEC side of
the resonance, we find the concave-down section of a curve moves toward
higher population imbalance and finally disappears in the $\cal EP$
landscape (shown in Fig. \ref{2} (d)). Homogeneous superfluids with
population imbalance are always stable in this limit.

Before concluding we would like to make two comments. The first one is
about the role of population imbalance relaxation (PIR). In the
presence of PIR, the population imbalance takes a unique equilibrium
value $P^{eq}_{im}(I)$; phase separation of cold atoms in
this limit (with the total number density fixed) needs a fine tuning in
the
energy splitting $I$. If the difference between the density
of a BCS state and that of a normal state at the same chemical potential 
is negligible, the phase separation only occurs when $I$ takes a critical value. 
Indeed, as
illustrated before\cite{Sheehy} the phase separation takes place in the
vicinity of a critical splitting and the interval becomes to be
exponentially small in the weakly interacting limit.
In the view of this, the phase separation observed in
Ref.\cite{Zwierlein,Partridge}
in the {\em extreme quantum limit} (i.e. extremely slow imbalance
relaxation) is
distinct from phase separation phenomena 
in the opposite limit where only equilibrium states are considered.
Meanwhile, in general the time-dependence of population imbalance
can be expressed as 
$P_{im}(t)=P_{im}(0)\exp(-t/\tau)+P^{eq}_{im}(I)(1-\exp(-t/\tau))$ (here 
$\tau$ is the imbalance relaxation time.). 
The temporal behavior of phase separation therefore does 
depend on the energy splitting $I$, though practically 
such dependence is always suppressed if the
measurement time is much shorter than the relaxation time $\tau$.

The second remark is on the effect of confinement or finite size. In
the presence of smooth potentials, the spatial variation of particle
density effectively provides an ensemble of cold atoms superfluids
at different chemical potentials. The phase separation occurs in a
region where the self-consistent chemical potential difference 
$\delta \mu $ (determined
by the population imbalance) is equal to the critical value $\delta \mu_c$ 
for the local self-consistent chemical potential $\mu(\bf r)$. 
However, if the confinement potential is
hard-wall like, then the chemical potential difference
could be {\em self-consistently} pinned in the vicinity of 
the critical value $\delta \mu_c$ for the bulk density.

In conclusion, we have mapped out the $\cal EP$ landscape for fermion
states with population imbalance. This result might be the blueprint
for the future studies of the dynamics of various competing states near
Feshbach resonances. For instance it can be used to understand the
energy released, during the phase separation,
from a homogeneous state initially prepared and the evolution of 
initial states. The phase separation of fermion pairing states
in a population imbalance conserved subspace turns out to be robust and needs no fine
tuning in the energy splitting.
This is also the limit where experiments were carried out
and we believe the results obtained here will help to identify new
interesting experimental opportunities in the future.  We would like to thank Mike Forbes, Randy Hulet, Wolfgang
Ketterle, Tony Leggett, and Boris Spivak for numerous discussions. 
This work is supported by a grant from the office of the Dean of Science at University of British Columbia and
a Discovery grant from NSERC, Canada. FZ is an A. P. Sloan
fellow.


\begin{thebibliography}{10}
\bibitem{Zwierlein} M. W. Zwierlein, et al., Science \textbf{311}, 
492 (2006).

\bibitem{Partridge} G. B. Partridge, et al., Science \textbf{311}, 
503 (2006). 

\bibitem{Clogston} M. Clogston, Phys. Rev. Lett. \textbf{9}, 266 (1962).

\bibitem{Sarma} G. Sarma, J. Phys. Chem. Solids \textbf{24}, 1029 (1963).

\bibitem{Fulde} P. Fulde and R. A. Ferrell, Phys. Rev. \textbf{135}, 
A550 (1964).

\bibitem{Larkin} A. J. Larkin and Y. N. Ovchinnokov, 
Sov. Phys. JETP \textbf{20}, 762 (1965).

\bibitem{Liu} 
W. V. Liu and F. Wilczek, Phys. Rev. Lett. 
\textbf{90}, 047002 (2003); M. Forbes et al., Phys. Rev. Lett. 
\textbf{94}, 017001 (2005); T. L. Ho and H. Zhai, cond-mat/0602568.


\bibitem{Bedaque} P. F. Bedaque, H. Caldas and G. Rupak, 
Phys. Rev. Lett. \textbf{91}, 247002 (2003).


\bibitem{exp1} M. Greiner, et al., Nature \textbf{426}, 537 (2003); C. A.
Regal, et al., Phys. Rev. Lett. \textbf{92}, 040403 (2004); S.
Jochim, et al., Science \textbf{302}, 2101 (2003); M. W. Zwierlein,
et al., Phys. Rev. Lett. \textbf{91}, 250401 (2003); K. E. Strecker,
et al., Phys. Rev. Lett. \textbf{91}, 080406 (2003).





\bibitem{Carlson} J. Carlson and S. Reddy, Phys. Rev. Lett. \textbf{95}, 060401 (2005).

\bibitem{Pao} 
C. H. Pao, et al., cond-mat/0506437;
D. T. Son and M. A. Stephanov, cond-mat/0507586.


\bibitem{Sheehy} D. E. Sheehy and L. Radzihovsky, Phys. Rev. Lett.
\textbf{96}, 060401 (2006).

\bibitem{Duan06}
P. Pieri and G. C. Strinati, cond-mat/0512354;
W. Yi and L. M. Duan, cond-mat/0601006;
F. Chevy, cond-mat/0601122;
T. N. De Silva and E. J. Mueller, cond-mat/0601314; M. Haque and
H. T. C. Stoof, cond-mat/0601321.




\end{thebibliography}
\end{document}